\documentclass[bibyear]{aa} 

\usepackage{graphicx}
\usepackage{amsmath}
\usepackage{float} 
\usepackage{txfonts}
\begin{document}

%
  \title{Limb and gravity-darkening coefficients for the {{\sc TESS}} 
satellite  at several metallicities, surface gravities, and 
microturbulent velocities }
 { }
   \subtitle{  }
\author{A. Claret}
   \offprints{A. Claret, e-mail:claret@iaa.es. Tables 2-29 are   available 
in electronic form at the CDS via anonymous ftp. }
\institute{Instituto de Astrof\'{\i}sica de Andaluc\'{\i}a, CSIC, Apartado 3004,
            18080 Granada, Spain}
            \date{Received; accepted; }
\abstract
   {}
   { We  present new  gravity and limb-darkening
coefficients for a wide range of effective temperatures, gravities, metallicities, and 
microturbulent velocities. These coefficients can be used in many different fields of 
stellar physics as synthetic light curves of eclipsing binaries and planetary transits, 
stellar diameters,   line profiles in rotating stars, and others. }
   {The limb-darkening coefficients were computed specifically for the photometric system 
of the space mission {{\sc TESS}} and were performed by  adopting the least-square method.  
In addition,  the linear and bi-parametric coefficients, by adopting the flux conservation 
method, are also available. On the other hand, to take into account  the effects of tidal 
and rotational distortions, we computed the passband gravity-darkening coefficients 
$y(\lambda)$ using a  general differential equation in which  we consider   the effects 
of convection and of the partial derivative 
$\left(\partial{\ln I(\lambda)}/{\partial{\ln g}}\right)_{T_{\rm eff}}$. }
   {To generate the limb-darkening coefficients we adopt two stellar atmosphere models: 
 {{\sc ATLAS}} (plane-parallel)  and {{\sc PHOENIX}} (spherical, quasi-spherical, and $r$-method). 
 The specific intensity distribution was fitted using five approaches: linear, 
quadratic, square root, logarithmic,   and a more general one with four terms. 
These grids cover together 19 metallicities ranging from 10$^{-5}$ up to 10$^{+1}$ solar 
abundances, 0 $\leq$  log g $\leq$ 6.0 and 1500 K $\leq$ T$_{\rm eff}$ 
$\leq$ 50000 K. The calculations of the gravity-darkening coefficients were performed 
for all plane-parallel {{\sc ATLAS}} models. 
 }
   {}

   \titlerunning {Limb-darkening coefficients }
   \maketitle
%

\section{Introduction}

The limb-darkening coefficients (LDC) are a fundamental tool in several 
areas of stellar physics, as for example: eclipsing binaries, measurement 
of stellar diameters,  line profiles in rotating stars, gravitational 
micro-lensing, optical interferometry, or more recently,  extra-solar 
planets' transits. Despite the  advances in the semi-empirical derivations 
of LDC, there is still   a serious scarcity of this kind of data.   
Because of this, we  are not yet able to perform a robust and consistent 
inter-comparison between observational and theoretical LDC. However, in the past few 
years an important effort has been carried out  in this direction.  
For details of these comparisons related to  eclipsing binaries and planetary 
transits see for example Claret (2008), Southworth (2008), Claret (2009), Sing (2010), 
Howarth (2011), Claret \& Bloemen (2011), Southworth (2012), Csizmadia et al. (2013), 
and M\"uller et al. (2013). 

One of the most important sources of the semi-empirical data of LDC are  the  space 
missions such as  {\it Kepler},  CoRoT,  and {\sc MOST} (Microvariability and Oscillations in STars) which are providing extensive 
observational material of high quality which is enabling a more comprehensive 
comparison.  Complementing this set of space missions, next year the satellite  
{\sc TESS} (Transiting Exoplanet Survey Satellite) is  scheduled to launch and the 
data collected by all these instruments will enable the researchers to extend even further    
the comparison of semi-empirical data with the theoretical LDC in some favourable cases. 
These comparisons are crucial  since they  may provide some clues to the  stellar 
atmosphere modellers which could help  to improve the theoretical models.
As a small contribution to this effort, we present in this paper the theoretical 
calculations of the limb and gravity-darkening coefficients (GDC)   for the 
photometric system of the satellite {\sc TESS}. 

The paper is organized as follows. In Section 2 we  describe briefly the  objectives 
and main characteristics of the {\sc TESS} mission. Section 3 is devoted to  introducing 
the numerical methods and the stellar atmosphere models used to compute the LDC and GDC, 
while in Section 4 we discuss the results.
   
\section {A brief summary of the space mission  {\sc TESS}}

Following the successful steps by the {\it Kepler} mission, {\sc TESS} is a new  
exo-planet finder to be launched by NASA in 2017  that will perform an all-sky survey.  
It is expected that this instrument will explore about   200,000 stars in the solar 
neighbourhood, searching for exo-planets through the planetary transit technique. 
The stars which are planned to be observed with {\sc TESS} are brighter, on average, 
than those observed by {\it Kepler} and  the spectral types to be surveyed range from 
 F5 to M5. On the other hand, the sky area to be covered is 400 times larger than that 
covered by  {\it Kepler}. The mentioned characteristics, among others, will permit 
the identification of a wide range of planets  from the size of Earth to gas giants, in diverse 
orbital configurations. Therefore, it will be possible to investigate some  fundamental 
planetary properties such as  mass, radius, orbital dynamics, and details on the atmosphere 
(for more information on the mission, see Ricker et al.  (2015), Sullivan et al. (2015) 
and references given therein). 

Within this context the calculations of the LDC are  important since they will allow 
the synthesis of light curves to be compared with the observational data coming from 
{\sc TESS}. As mentioned in the Introduction, these inter-comparisons may supply new and 
more accurate data for the semi-empirical LDC, providing a crucial test   to the stellar 
atmosphere models.

\section {Numerical methods and the  grids of stellar atmosphere models}

The  LDC laws adopted here are written down to facilitate the identification of 
the coefficients contained in the respective tables:   

\noindent
the linear law
\begin{eqnarray}
          \frac{I(\mu)}{ I(1)} = 1 - u(1 - \mu),
\end{eqnarray}

\noindent
the quadratic law

\begin{eqnarray}
        \frac{I(\mu)}{ I(1)} = 1 - a(1 - \mu) - b(1 - \mu)^2,
\end{eqnarray}

\noindent
the square root law

\begin{eqnarray}
      \frac{I(\mu)}{ I(1)} =  1 - c(1 - \mu) - d(1 - \sqrt{\mu}),
\end{eqnarray}

\noindent
the logarithmic law

\begin{eqnarray}
\frac{I(\mu)}{ I(1)} =  1 - e(1 - \mu) - f\mu\ln(\mu),
\end{eqnarray}

\noindent
and a four terms law introduced by us some time ago: 

\begin{eqnarray}
\frac{I(\mu)}{ I(1)} = 1 - \sum_{k=1}^{4} {a_k} (1 - \mu^{\frac{k}{2}}).
\end{eqnarray}

\noindent
In the above equations  $I(1)$ is the specific intensity at the centre of the disk and 
 $u, a, b, c, d, e, f$, and $a_k$ are the corresponding LDC. The  $\mu$'s are given by 
$\cos(\gamma)$, where $\gamma$ is the angle between the line of sight and  the outward 
surface normal. The model atmosphere intensities were convolved with  the  transmission 
curve for {\sc TESS},   provided by  D. Latham (2015, private communication). As in the 
previous papers in this series, we have used the ATLAS (plane-parallel geometry),  
 {\sc PHOENIX-COND} with spherical geometry  (Husser et al. 2013), and {\sc PHOENIX-DRIFT}  
also with spherical geometry (Witte et al. 2009). These grids cover together 19 metallicities 
ranging from 10$^{-5}$ up to 10$^{+1}$ solar abundance, 0 $\leq$  log g $\leq$ 6.0 and 
1500 K $\leq$ T$_{\rm eff}$ $\leq$ 50000 K. The values of the microturbulent  velocities 
($V_{\xi}$) are 0, 1, 2, 4, 8 km/s.

The LDC were computed adopting the least  square  method (LSM)  that allows a very good 
description   of I($\mu$) at any part of the disk, for any  log $g$, effective temperature, 
metallicity, and microturbulent velocity, mainly if  Eq. 5 is adopted. Indeed, it is 
desirable to handle an approach such as Eq. 5 that presents the following characteristics:  
a)  a single law which is valid for the whole HR (Hertzsprung-Russell) diagram, b)  is capable of reproducing 
well the intensity distribution and which as a consequence conserves the flux within a 
very small tolerance, c) can be applicable to different pass-bands, monochromatic 
quantities, chemical compositions, effective temperatures, gravities, and microturbulent 
velocities. For the case of LSM, we  adopted equally spaced $\mu$ points to prevent 
too large weights being applied to the limb.  For completeness, we also computed the LDC adopting  
the flux conservation method  (FCM),   for bi-parametric laws and for the  linear 
approximation; to differentiate, we used the suffixes LSM and FCM in the tables. 

The dispersion of the specific intensities associated to the FCM can be as large 
as 1000 times those provided by the LSM when Eq. 5 is adopted. This is a serious 
restriction to the FCM since a good match is necessary to compute the loss and gain of 
light during the eclipses. Moreover, if a non-linear law is used, an extra condition 
must be introduced, in addition to the flux conservation. This extra condition is 
often arbitrary.  On the other hand, the inter-comparison between the LDC computed 
with both numerical methods can serve as a tool to estimate the theoretical error
bars.  We refer interested readers to Claret (2000) for a more detailed 
discussion on this subject.

Due to the  different input physics and numerical resolutions, the {\sc PHOENIX} models 
are divided into two sets:  1500 K $\leq$ T$_{\rm eff}$ $\leq$ 3000 K ({\sc DRIFT}) and 
another  with 2300 K $\leq$  T$_{\rm eff}$ $\leq$ 12000 K ({\sc COND}). Still concerning the 
{\sc PHOENIX} models, we have introduced in past papers the concept of quasi-spherical 
models for main sequence stars. For this kind of stars spherical models are constituted by a core  ($\mu > $ 0.1) 
that behaves  like a plane-parallel structure, and by an envelope that delivers the 
spherical part of the intensities ($\mu \lesssim$ 0.1. When we compare the intensity 
distribution of a spherical model with one  adopting the plane-parallel  geometry but 
with the same input physics (log $g$, metallicity, T$_{\rm eff}$, $V_{\xi}$), we detect 
a  similar intensity distribution for both models, except in the drop-off region 
($\mu \lesssim$  0.1). We  define a quasi-spherical model  as the model computed 
adopting the  spherical symmetry but whose LDC are computed only for the core, that is, 
without considering the points inside the drop-off region. This  concept  allows us  
to  compare the LDC for {\sc PHOENIX} and {\sc ATLAS} models (see  below). It can be also  
used in situations where the effects of sphericity are not important.

Before discussing the results of the  LDC for {\sc TESS}, it is convenient to investigate 
an alternative to the usual quasi-spherical models. Based on the work by  Wittkowskii,
Aufdenberg, \& Kervella (2004), Espinoza \& Jordan (2015) used   $r = \sqrt{(1-\mu^2)}$,  
instead of $\mu$. They derive the LDC by searching for the maximum of the derivative of
the specific intensity with respect to $r$ and shifting the profile to this point 
(hereafter referred as   $r$-method).  These  authors have found large differences when 
comparing the quadratic LDC ({\it Kepler}) computed using the $r$-method with the quadratic LDC  using 
quasi-spherical models by Claret, Hauschildt, \& Witte (2012), mainly for cooler models as
 seen in their Fig. 6. However, this comparison was not performed  for the same {\sc PHOENIX} 
models: Claret, Hauschildt, \&  Witte used the  {\sc PHOENIX-DRIFT} models for effective 
temperatures lower than 3000 K, while Espinoza \& Jordan  adopted the {\sc PHOENIX-COND} ones 
(Espinoza, private communication). This important point  was not considered by the mentioned 
authors in that comparison.  The differences between  {\sc PHOENIX-DRIFT} and {\sc PHOENIX-COND} 
models are large, as  discussed  in Claret, Hauschildt, \& Witte (Sect. 2). Moreover, the 
LDC computed  for larger effective temperatures by  Claret, Hauschildt, \& Witte are also not completely 
suitable for direct comparison  in Fig. 6 by  Espinoza \& Jordan,   because they also  come 
from different stellar atmosphere models from those adopted by these authors.

To try to clarify this point we  compute quadratic LDC ({\it Kepler})  for {\sc PHOENIX-COND}  
models adopting the  usual quasi-spherical ($\mu > $ 0.1) and the $r$-method. This allows us 
to compare the LDC varying only the numerical methods for {\sc PHOENIX} models while adopting 
the same input physics.  Figure 1 shows such a comparison for the LDC as a function of the 
effective temperature, where  crosses and asterisks represent the calculations adopting the 
usual quasi-spherical  and $r$-method, respectively. The {\sc PHOENIX-DRIFT} (x) models are also
shown for comparison.  The differences between the LDC for the  {\sc PHOENIX-DRIFT} and 
{\sc PHOENIX-COND} models are very similar and  are of the same order as those found  by 
Espinoza \& Jordan in their Fig. 6.  However,  the total magnitude of these  differences 
in this case (log g =4.5) is not fully assessed due to the different adopted  methods 
(quasi-spherical and $r$) as Espinoza \& Jordan argued; these differences  are mainly related to the 
comparison between two different versions of  {\sc PHOENIX} models. A direct comparison 
between the LDC for  {\sc PHOENIX-COND} models adopting the usual quasi-spherical and 
$r$ methods shows that both procedures provide very similar LDC (Fig. 1) presenting only 
small differences in the  region 3.43 $<$ log T$_{\rm eff} < $ 3.65. These differences  
are probably related in part to the goodness of the fitting since in this interval the 
$r$-method provides the worse fittings (see third panel).

On the other hand,  numerical experiments show that the goodness of fittings also depends 
on the local gravity (drop-off), being in general  higher for the usual quasi-spherical method in the 
case of large log g. The opposite occurs for the $r$-method,  at least for  {\it Kepler} 
quadratic LDC. For the general case of bi-parametric laws, the differences between the LDC 
obtained using  the usual quasi-spherical and $r$ methods increase for small values of 
log g;  these differences  are, however,  very small for main sequence models, that is 
for log g $\geq$ 4.0. In the case of the linear law, the mentioned differences are small 
and almost independent of logg. To try to minimize  this problem, we redefine a 
quasi-spherical model  as before but instead of considering the points  $\mu >$ 0.1 we 
apply a simple algorithm to the normalized specific intensities to properly consider the 
exclusion of the drop-off region. We consider for  LDC calculations the $\mu$ points 
between 1.0 and the point where the normalized intensity decays to $0.1$.  This method has 
the advantage of preserving the original core points of the {\sc PHOENIX} models to effects 
of comparison with the other procedures. This new definition does not affect seriously the previous LDC 
calculations, particularly for main sequence models. Figure 1 confirms this point.  
Hereafter (including Table 1)  the quasi-spherical  models refer to the new definition.

Despite the  problems with the comparison of LDC performed by Espinoza \& Jordan discussed 
previously  and the similarities between the results from the  new quasi-spherical  and 
$r$ methods  mainly for main sequence models, the LDC for {\sc PHOENIX-DRIFT} and 
{\sc PHOENIX-COND} models are available by adopting both methods (Table 1). We also provide 
the respective merit function to guide the readers when selecting the more suitable LDC. 
The LDC   adopting  {\sc PHOENIX} models for other chemical compositions can be provided to 
the interested readers upon request.

\section {Discussion of the results}

\subsection {Limb-darkening}

 As  discussed extensively in the earlier papers on limb-darkening, the linear law  is not 
a suitable approximation for most of  the specific intensity distributions.  This non-linearity  
is even more conspicuous in the case of spherical models due to the drop-off  caused by the 
decreased matter-radiation interaction for $\mu \lesssim$ 0.1. Although this law presents 
this problem, it can still be useful for  comparing models with  different geometries and/or 
input physics. For example,  in Fig. 2 we plot the linear LDC({\sc TESS})  for {\bf{\sc PHOENIX-COND}} 
(quasi-spherical)  and {\sc ATLAS} models, with log[A/H = 0.0,  $V_{\xi}$ = 2 km/s and log~$g$ = 4.5. 
The continuous line denotes the {\sc ATLAS} models while the dashed line indicates the LDC for   
{\sc PHOENIX-COND} models. The agreement can be considered as good, except in  the onset of 
convection,  because both sets of models  were computed with different  mixing-length parameters. 
Also, for lower effective temperatures the agreement is not so good, due to the different  
input physics for cooler models.

   \begin{figure}
   \includegraphics[height=8.cm,width=6.4cm,angle=-90]{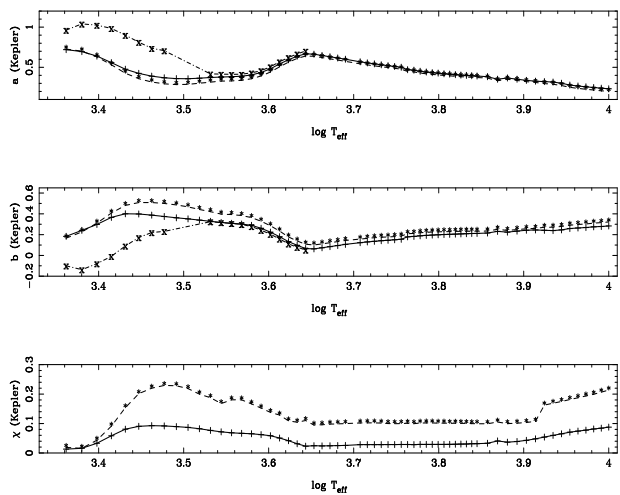}

   \caption{ Quadratic LDC for {\sc PHOENIX-COND} models. Quasi-spherical models are  represented by a continuous line and crosses  and   the $r$ method by a dashed line and asterisks.  
The LDC for the {\sc PHOENIX-DRIFT} models from Claret, Hauschildt, \& Witte  (2012) 
are denoted by a dashed-dotted line and the symbol x. In the last case and for the 
sake of clarity,  only models with T$_{\rm eff}$ $<$ 4400 K are shown. The third panel 
shows the root square of the merit function $\chi$.  All calculations were performed 
for log $g$ = 4.5, log[A/H]= 0.0, $V_{\xi}$ = 2 km/s. {\it Kepler} photometric system. 
LSM calculations. }
   \end{figure}

   \begin{figure}
   \includegraphics[height=8.cm,width=6.4cm,angle=-90]{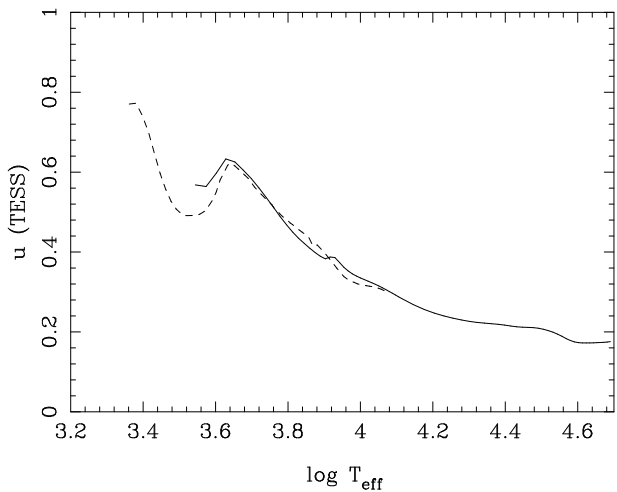}
   \caption{Theoretical linear LDC for  {\sc ATLAS} models (continuous line) and 
{\sc PHOENIX-COND} quasi-spherical ones (dashed line). Log $g$ = 4.5, log[A/H]= 0.0, 
$V_{\xi}$ = 2 km/s. {\sc TESS} photometric system. LSM calculations. }
   \end{figure}

   \begin{figure}
   \includegraphics[height=8.cm,width=6.4cm,angle=-90]{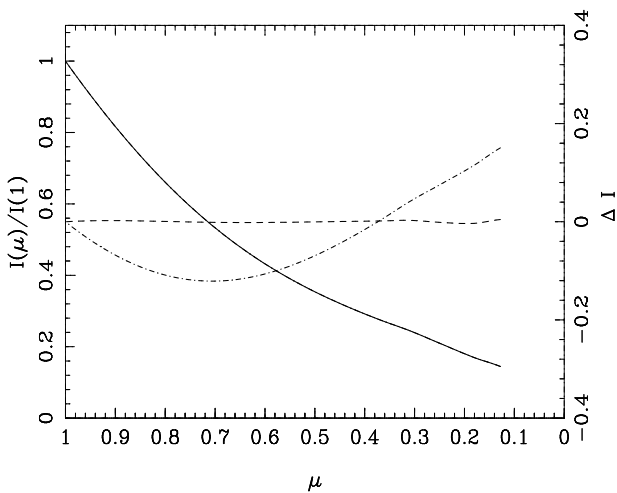}
   \caption{Specific intensity distribution for a [1500 K, 4.5] {\sc PHOENIX-DRIFT} 
quasi-spherical symmetric  model. Continuous line represents the model intensities 
(left label), while the dashed line denotes the deviations [model-fit] by adopting Eq. 5  
and the dashed-dotted one represents the deviations by adopting Eq. 1 (right label). 
Log[A/H]= 0.0, $V_{\xi}$ = 2 km/s. {\sc TESS} photometric system. LSM calculations.    }
   \end{figure}

 \begin{figure}
   \includegraphics[height=8.cm,width=6.4cm,angle=-90]{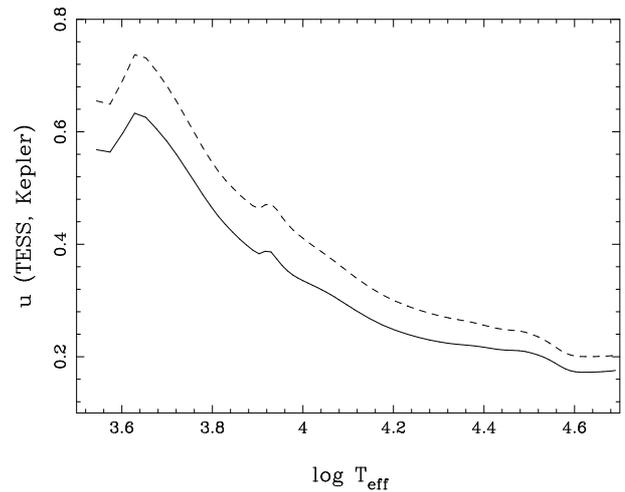}
   \caption{Theoretical linear LDC for  {\sc ATLAS} models.  Continuous line denotes  
 {\sc TESS} while dashed line represents the {\it Kepler} photometric system. Log $g$ = 4.5, 
log[A/H]= 0.0, $V_{\xi}$ = 2 km/s. LSM calculations.  }
   \end{figure}

\begin{figure}
   \includegraphics[height=8.cm,width=6.4cm,angle=-90]{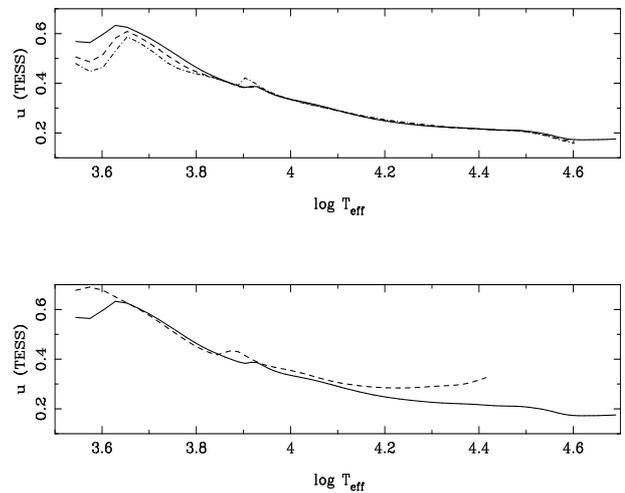}
   \caption{Effects of metallicity and evolutionary status on the theoretical linear 
LDC for  {\sc ATLAS} models, {\sc TESS} photometric system.  Upper panel:  continuous 
line denotes models with  log[A/H]= 0.0 while  dashed line indicates log[A/H]= -0.5 and  
dashed-dotted line those with log[A/H]= -1.0. Log $g$ = 4.5 and $V_{\xi}$ = 2 km/s 
for all models. Lower panel:  continuous line denotes models with  log $g$= 4.5 and 
dashed line represents models with  log $g$= 3.0. Log [A/H] = 0.0 and $V_{\xi}$ = 2 km/s 
for all models. LSM calculations for both panels.  }
   \end{figure}

   \begin{figure}
   \includegraphics[height=8.cm,width=6.4cm,angle=-90]{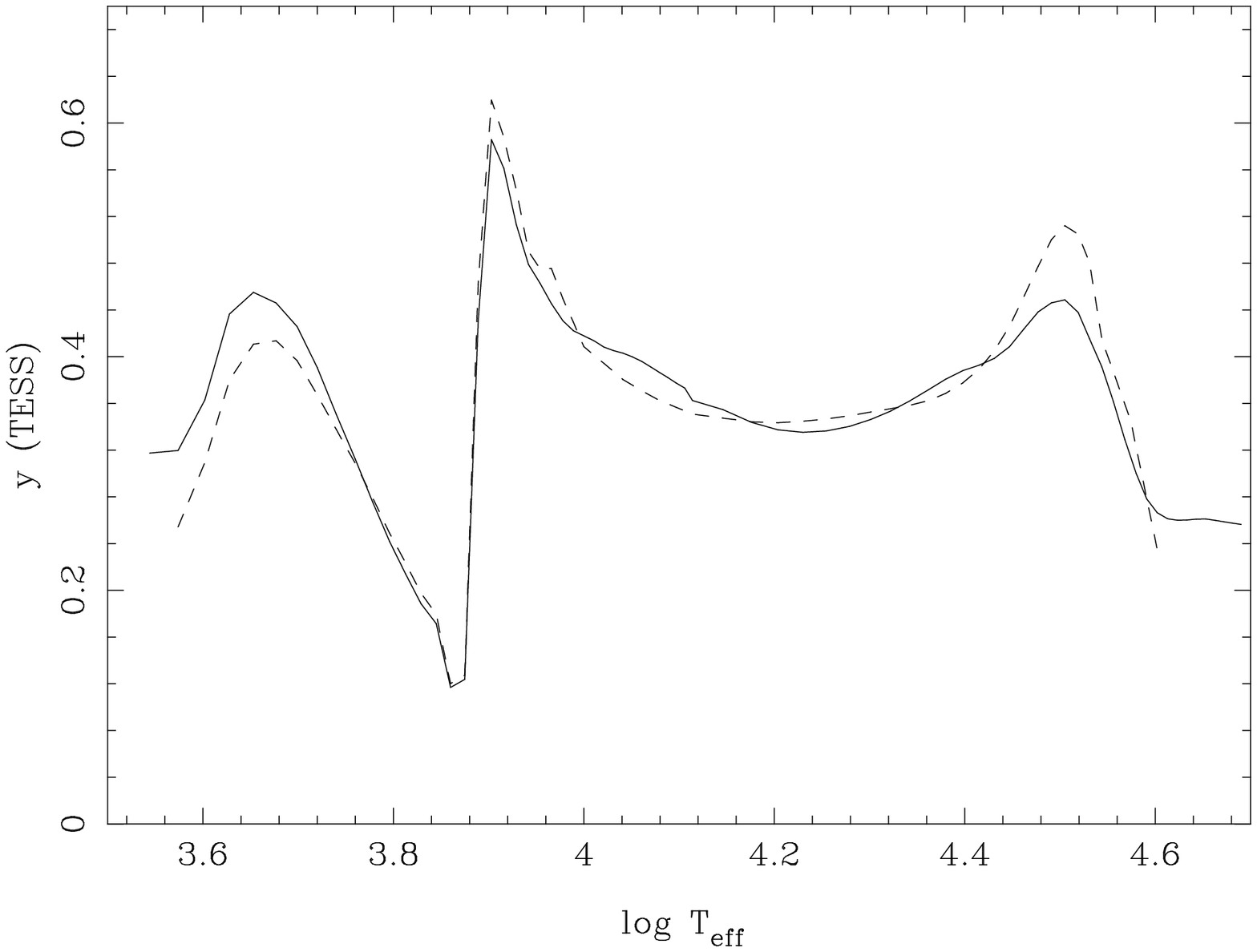}
   \caption{Theoretical gravity-darkening coefficients for  {\sc TESS}, {\sc ATLAS} 
models. The continuous line represents the calculations for log [A/H] = 0.00 and  and 
dashed line is for  log [A/H] = -2.5.  Log $g$ = 4.5, log[A/H]= 0.0, $V_{\xi}$ = 2 km/s.}
   \end{figure}

It is worth noticing the high values of  the linear LDC (larger than 1.0) for  the  
{\sc PHOENIX-DRIFT} models at lowest effective temperatures, since it implies a negative relative 
intensity at the limb. This is a consequence of the not so good fit provided by Eq. 1.  In 
Fig. 3  we show the behaviour of a {\sc PHOENIX} model with  T$_{\rm eff}$ = 1500 K, log[A/H] = 0.0,  
$V_{\xi}$ = 2 km/s, and log $g$ = 4.5. The model intensity is represented by a continuous  line while 
the dashed one denotes the deviation of the intensity [model-fitted] using Eq. 5 and the dashed-dotted 
line represents the deviation using Eq. 1. As a consequence of this, we recommend using the approach 
given by  Eq. 5 or  at least a bi-parametric law for the lowest effective temperatures. 

A point that can be of special interest for observers is the behaviour of the LDC for the photometric 
systems of {\bf{\it Kepler}} and {\sc TESS}. Again we use the linear coefficient 
for the sake of clarity in the comparison. As the effective wavelengths 
are very different ($\approx$ 630 and $\approx$ 800 nm, respectively) the corresponding LDC follow the general 
trend with the effective wavelength being smaller for higher $\lambda_{\rm eff}$. The differences shown 
in Fig. 4 could be detected  within  the expected accuracy of semi-empirical LDC. Therefore, it will 
be very useful in the future  if observers compare their respective semi-empirical LDC obtained 
with  both instruments for similar target stars. It would also be very interesting to compare 
the empirical LDC with those  generated using the  two atmospheres' models adopted here, given that both 
cover the spectral range of {\sc TESS}.  

Other points of interest are the metallicity and evolutionary effects on the LDC. In Fig. 5 we 
can see a comparison between models computed with the solar abundance and less metallic ones 
(upper panel, log [A/H] = -0.5 and -1.0). The differences are more pronounced for models 
located after the onset of convection, that is, for log T$_{\rm eff} < $ 3.9. The influence of 
the evolutionary status is shown in the lower panel. The corresponding differences are large 
for the extreme ranges of effective temperatures and could be detected observationally. 

\subsection {Gravity-darkening}

 In a binary system the tides tend to elongate the stars along the line that joins them,  
while rotation tends to flatten them at the poles. The deviations from the spherical symmetry 
can be written as a function of the rotational rate and of the mass ratio. In addition to 
the geometrical perturbations  due to the proximity effects, there is also an associated  
thermodynamic change. In 1924,  von Zeipel established that the flux distribution in a 
distorted star is not uniform and is proportional   to $g^{\beta_1}$, where $g$ is the 
local gravity and $\beta_1$ is  the gravity-darkening exponent (GDE), usually taken as 
1.0 for radiative envelopes. This   $\beta_1$  is valid only  for stars in strict radiative 
equilibrium and when the diffusion equation is used as a transfer equation. Recently, 
Claret (2015, 2016) showed that, under determined physical conditions, the theorem by 
von Zeipel is no longer valid. 

The GDE is a bolometric quantity but observations are generally performed in photometric 
bands. In 1959 Kopal introduced the concept of gravity-darkening coefficients 
(GDC, $y(\lambda)$) which connect both the bolometric and pass-band quantities.  This concept 
is very useful to compute the light distribution of  distorted configurations. If we expand in a 
series the ratio between the monochromatic and total radiation,  we obtain the corresponding 
$y(\lambda)$. For simplicity, it was assumed that the distorted configurations radiate as a 
black-body, which is not   a good approximation. Often the following expression by Martynov 
(1973)  is adopted to compute the GDC:

\begin{eqnarray} 
y(\lambda, T_{\rm eff}, \log [A/H], \log g, V_{\xi}) = \frac{1}{4} 
\left(\frac{\partial\ln I(\lambda)} {\partial\ln T_{\rm eff}}\right)_{g}.    
\end{eqnarray}

In the above equation, $\lambda$ denotes the wavelength and $I$ the intensity at a given 
wavelength (or pass-band) at the centre of the stellar disc. This equation was improved later 
by Claret \& Bloemen (2011) to take into account the term  
$\left(\partial{\ln I(\lambda)}/{\partial{\ln g}}\right)_{T_{\rm eff}}$ 
and the effects of convection on $\beta_1$. Moreover, instead of the black-body approach we 
use the same {\sc ATLAS} atmosphere models used to compute the LDC.  Therefore, the general equation 
which will be adopted here to compute $y(\lambda)$ is:

\begin{eqnarray}
\lefteqn{y(\lambda, T_{\rm eff }, \log [A/H], \log g, V_{\xi}) =} \nonumber \\ 
\lefteqn{\left(\frac{d\ln T_{\rm eff }}{{d\ln g}}\right)
\left(\frac{\partial{\ln I(\lambda)}}{{\partial{\ln T_{\rm eff }}}}\right)_{g}
 + \left(\frac{\partial{\ln I(\lambda)}}
{\partial{\ln g}}\right)_{T_{\rm eff}}.}
\end{eqnarray}

The effects of the convection on $y(\lambda)$ are important  for cooler models and 
 $\left(d\ln T_{\rm eff }/{d\ln g}\right)$ was  computed   considering the bolometric 
GDE previously calculated (see Claret (2004)). For the models located in the main-sequence, 
the contribution of the partial derivative at constant effective temperature is not 
important,   but it is not negligible  for cool giants stars. Of course, Eq. 7 reduces to 
Eq. 6 if $\beta_1$ = 1.0 and the partial derivative at constant effective temperature are 
assumed to be zero. 

The results for the {\sc TESS} 
photometric system are shown in Fig. 6 for log [A/H] = 0.0 and -2.5. The effects of 
metallicity on $y(\lambda)$ are not very large, except for two regions centred in 
log T$_{\rm eff}$ = 3.7 and 4.5. The drop-off in $y(\lambda)$ around log T$_{\rm eff}$ = 
3.9 is due to the effects of convection on $\beta_1$. These computations supersede the 
old values of $y(\lambda)$  based on the black-body approximation. Due to the narrowness  
of the basic physics input (log T$_{\rm eff}$, log $g$, log [A/H], $V_{\xi}$)  of the 
{\sc PHOENIX} grids, we have computed $y(\lambda)$ only for the {\sc ATLAS} models. However, 
the  calculations adopting  {\sc PHOENIX} models can be provided to interested readers 
upon request. Finally, Table 1 summarizes the kind of data available  at CDS (Centre de Donn\'ees Astronomiques de Strasbourg)  or directly  from the author. 

\begin{acknowledgements} 
I thank the anonymous referee for his or her comments on the   original manuscript. 
The author would like to thank D. Latham and G. Torres for suggesting this work, 
for providing the transmission curve of TESS, and for helpful discussions.  
I acknowledge  T.-O. Husser and  P. Hauschildt  for providing the PHOENIX models. 
I also would like to thank  B. Rufino for her comments.  The Spanish MEC 
(AYA2015-71718-R and ESP2015-65712-C5-5-R) is gratefully  acknowledged for its 
support during the development of this work. This research has made use of the 
SIMBAD database, operated at the CDS, Strasbourg, France, and of NASA's Astrophysics 
Data System Abstract Service.
\end{acknowledgements}

{}

\begin{table}
\caption[]{Gravity and limb-darkening coefficients for the {\sc TESS}  photometric system}
\begin{flushleft}
\begin{tabular}{lccccccl}                         
\hline                         
Name    & Source   &  range T$_{\rm eff}$ & range log $g$ & log [A/H] & Vel Turb. & Filter & Fit/equation/model   \\ 
\hline   
Table2  &{\sc PHOENIX-DRIFT}& 1500 K-3000 K & 2.5-5.5&  0.0  & 2 km/s&{\sc TESS} & LSM/FCM/Eq. 1 quasi-spherical\\
Table3  &{\sc PHOENIX-DRIFT}& 1500 K-3000 K & 2.5-5.5&  0.0  & 2 km/s&{\sc TESS} & LSM/FCM/Eq. 1 $r$\\
Table4  &{\sc PHOENIX-DRIFT}& 1500 K-3000 K & 2.5-5.5&  0.0  & 2 km/s&{\sc TESS} & LSM/FCM/Eq. 2 quasi-spherical\\
Table5  &{\sc PHOENIX-DRIFT}& 1500 K-3000 K & 2.5-5.5&  0.0  & 2 km/s&{\sc TESS} & LSM/FCM/Eq. 2 $r$\\
Table6  &{\sc PHOENIX-DRIFT}& 1500 K-3000 K & 2.5-5.5&  0.0  & 2 km/s&{\sc TESS} & LSM/FCM/Eq. 3 quasi-spherical\\
Table7  &{\sc PHOENIX-DRIFT}& 1500 K-3000 K & 2.5-5.5&  0.0  & 2 km/s&{\sc TESS} & LSM/FCM/Eq. 3 $r$\\
Table8  &{\sc PHOENIX-DRIFT}& 1500 K-3000 K & 2.5-5.5&  0.0  & 2 km/s&{\sc TESS} & LSM/FCM/Eq. 4 quasi-spherical\\
Table9  &{\sc PHOENIX-DRIFT}& 1500 K-3000 K & 2.5-5.5&  0.0  & 2 km/s&{\sc TESS} & LSM/FCM/Eq. 4 $r$\\
Table10 &{\sc PHOENIX-DRIFT}& 1500 K-3000 K & 2.5-5.5&  0.0  & 2 km/s&{\sc TESS} & LSM/FCM/Eq. 5 quasi-spherical\\
Table11 &{\sc PHOENIX-DRIFT}& 1500 K-3000 K & 2.5-5.5&  0.0  & 2 km/s&{\sc TESS} & LSM/FCM/Eq. 5 $r$\\
Table12 &{\sc PHOENIX-DRIFT}& 1500 K-3000 K & 2.5-5.5&  0.0  & 2 km/s&{\sc TESS}  & LSM/Eq. 5     spherical\\
Table13  &{\sc PHOENIX-COND}& 2300 K-12000 K & 2.5-6.0&  0.0  & 2 km/s&{\sc TESS} & LSM/FCM/Eq. 1 quasi-spherical\\
Table14  &{\sc PHOENIX-COND}& 2300 K-12000 K & 2.5-6.0&  0.0  & 2 km/s&{\sc TESS} & LSM/FCM/Eq. 1 $r$\\
Table15  &{\sc PHOENIX-COND}& 2300 K-12000 K & 2.5-6.0&  0.0  & 2 km/s&{\sc TESS} & LSM/FCM/Eq. 2 quasi-spherical\\
Table16  &{\sc PHOENIX-COND}& 2300 K-12000 K & 2.5-6.0&  0.0  & 2 km/s&{\sc TESS} & LSM/FCM/Eq. 2 $r$\\
Table17  &{\sc PHOENIX-COND}& 2300 K-12000 K & 2.5-6.0&  0.0  & 2 km/s&{\sc TESS} & LSM/FCM/Eq. 3 quasi-spherical\\
Table18  &{\sc PHOENIX-COND}& 2300 K-12000 K & 2.5-6.0&  0.0  & 2 km/s&{\sc TESS} & LSM/FCM/Eq. 3 $r$\\
Table19  &{\sc PHOENIX-COND}& 2300 K-12000 K & 2.5-6.0&  0.0  & 2 km/s&{\sc TESS} & LSM/FCM/Eq. 4 quasi-spherical\\
Table20  &{\sc PHOENIX-COND}& 2300 K-12000 K & 2.5-6.0&  0.0  & 2 km/s&{\sc TESS} & LSM/FCM/Eq. 4 $r$\\
Table21 &{\sc PHOENIX-COND}& 2300 K-12000 K & 2.5-6.0&  0.0  & 2 km/s&{\sc TESS} & LSM/FCM/Eq. 5 quasi-spherical\\
Table22 &{\sc PHOENIX-COND}& 2300 K-12000 K & 2.5-6.0&  0.0  & 2 km/s&{\sc TESS} & LSM/FCM/Eq. 5 $r$\\
Table23 &{\sc PHOENIX-COND}& 2300 K-12000 K & 2.5-6.0&  0.0  & 2 km/s&{\sc TESS} & LSM/Eq. 5 spherical\\
Table24  &{\sc ATLAS}&  3500 K-50000 K & 0.0-5.0& -5.0-+1.0 & 0-8 km/s&{\sc TESS} & LSM/FCM/Eq. 1 \\
Table25  &{\sc ATLAS}&  3500 K-50000 K & 0.0-5.0& -5.0-+1.0 & 0-8 km/s&{\sc TESS} & LSM/FCM/Eq. 2 \\
Table26  &{\sc ATLAS}&  3500 K-50000 K & 0.0-5.0& -5.0-+1.0 & 0-8 km/s&{\sc TESS} & LSM/FCM/Eq. 3 \\
Table27  &{\sc ATLAS}&  3500 K-50000 K & 0.0-5.0& -5.0-+1.0 & 0-8 km/s&{\sc TESS} & LSM/FCM/Eq. 4 \\
Table28  &{\sc ATLAS}&  3500 K-50000 K & 0.0-5.0& -5.0-+1.0 & 0-8 km/s&{\sc TESS} & LSM/Eq. 5     \\
Table29  &{\sc ATLAS}&  3500 K-50000 K & 0.0-5.0& -5.0-+1.0 & 0-8 km/s&{\sc TESS} &  Gravity-darkening coefficients $y(\lambda)$    \\
\hline
\hline
\end{tabular}
\end{flushleft}
\end{table}

\end{document}